\documentclass[conference]{IEEEtran}
\IEEEoverridecommandlockouts
\usepackage{amsmath,amssymb,amsfonts}
\usepackage{algorithm}
\usepackage{algorithmic}
\usepackage{array}
\usepackage[caption=false,font=normalsize,labelfont=sf,textfont=sf]{subfig}
\usepackage[group-separator={,},group-minimum-digits=3]{siunitx}
\usepackage{textcomp}
\usepackage{stfloats}
\usepackage{url}
\usepackage{verbatim}
\usepackage{graphicx}
\usepackage{balance}
\usepackage{mathrsfs}
\usepackage{cite}
\usepackage{xcolor}
\usepackage{booktabs}

\usepackage[linewidth=1pt]{mdframed}
\usepackage[colorlinks=true,
            linkcolor=blue,
            anchorcolor=blue,
            citecolor=blue]{hyperref}
\begin{document}

\title{ 
{IRS-Enhanced Anti-Jamming Precoding Against DISCO Physical Layer Jamming Attacks}
}
\author{\IEEEauthorblockN{Huan~Huang\IEEEauthorrefmark{1}, Hongliang~Zhang\IEEEauthorrefmark{2}, Yi~Cai\IEEEauthorrefmark{1}, Yunjing~Zhang\IEEEauthorrefmark{1}, A.~Lee~Swindlehurst\IEEEauthorrefmark{3}, and~Zhu~Han\IEEEauthorrefmark{4}}
\IEEEauthorblockA{\IEEEauthorrefmark{1} Electronic Information School, Soochow University, Suzhou, China}
\IEEEauthorblockA{\IEEEauthorrefmark{2} State Key Laboratory of Advanced Optical Communication Systems and Networks, \\School of Electronics, Peking University, Beijing, China}
\IEEEauthorblockA{\IEEEauthorrefmark{3} Department of Electrical and Computer Engineering, University of Houston, Houston, USA}
\IEEEauthorblockA{\IEEEauthorrefmark{4} Center for Pervasive Communications and Computing, University of California, Irvine, USA}
}
\maketitle

\begin{abstract}
Illegitimate intelligent reflective surfaces (IRSs) can pose significant physical layer security risks on multi-user multiple-input single-output (MU-MISO) systems.
Recently, a DISCO approach has been proposed an illegitimate IRS with random and time-varying reflection coefficients, 
referred to as a ``disco" IRS (DIRS). 
Such DIRS can attack MU-MISO systems without relying on 
either jamming power or channel state information (CSI), and classical anti-jamming techniques are ineffective for
the DIRS-based fully-passive jammers (DIRS-based FPJs). 
In this paper, we propose an  IRS-enhanced anti-jamming precoder
against DIRS-based FPJs that requires only statistical rather than instantaneous CSI of the DIRS-jammed channels. 
Specifically, a legitimate IRS is introduced to reduce the strength of the DIRS-based jamming relative to the transmit signals at a legitimate user (LU).
In addition,  the active beamforming at the legitimate access point (AP) is designed 
to maximize the signal-to-jamming-plus-noise ratios (SJNRs).
Numerical results are presented to evaluate
the effectiveness of the proposed IRS-enhanced anti-jamming precoder against DIRS-based FPJs.
\end{abstract}


\section{Introduction}\label{Intro}
Due to the broadcast and superposition properties of wireless channels, 
wireless communications are vulnerable to malicious attacks 
such as physical layer jamming~\cite{PLSsur1,MyWCMag,AntiJammingSurv}.
Traditional active jammers (AJs) broadcast jamming signals, such as pseudorandom noise or modulated
Gaussian noise, to intentionally disrupt legitimate wireless communications like
Wi-Fi, Bluetooth, the Internet of Things (IoT), and cellular networks~\cite{PLSsur1,MyWCMag,AntiJammingSurv}.
Fortunately, energy constraints are an inherent drawback of AJs, severely limiting their use.
Moreover, classical anti-jamming approaches~\cite{AntiJammingSurv}, such as spread spectrum 
and frequency-hopping techniques can be used to effectively relieve the jamming impact imposed by AJs.

Recently, the interesting idea of disco intelligent reflective surfaces (DIRSs) 
has been reported~\cite{MyWCMag,DIRSVT,DIRSLS,DIRSTWC,MyGC23,MyGC23Extension},
where illegitimate IRSs with time-varying random reflection coefficients act like
``disco balls" to actively age wireless channels. 
Such active channel aging (ACA) can be used to jam legitimate users (LUs)
without either jamming power or LU channel state information (CSI). 
An IRS configured in this way is referred to as a DIRS-based fully-passive jammer (FPJ).
Moreover, some works have also investigated the use of DIRSs to  
break key consistency in channel reciprocity-based key generation~\cite{OtherDIRS1,OtherDIRS2,OtherDIRS3}. 
Some unique properties of the DIRS-based FPJs are that their jamming impact cannot 
be mitigated by increasing transmit power~\cite{DIRSTWC} and classical anti-jamming
approaches~\cite{AntiJammingSurv}, such as spread spectrum 
and frequency-hopping techniques, are not effective for DIRS-based FPJs. %

To address the serious concerns raised by DIRS-based FPJs, the previous work in~\cite{MyGC23, MyGC23Extension} provided a possible approach. 
This work showed that the elements of the DIRS-jammed channels
converge to a complex Gaussian distribution as the number of the DIRS reflective elements 
becomes large, and then designed  an anti-jamming precoder for the DIRS-based FPJs 
that requires only statistical rather than instantaneous CSI.
Furthermore, an approach to estimate the statistical characteristics of the DIRS-based
channels was proposed in~\cite{MyGC23Extension}.
However, this anti-jamming precoder cannot achieve good jamming suppression
at high transmit power~\cite{MyGC23, MyGC23Extension}.

The investigation conducted in~\cite{MyWCMag} suggested that the
legitimate AP can decrease the amount of DIRS-based
ACA interference relative to the strength of the transmit signals to
enhance the effect of  the anti-jamming precoder~\cite{MyGC23,MyGC23Extension}.
Based on this heuristic investigation, one can
introduce legitimate IRSs to weaken the strength of the DIRS-based ACA interference 
relative to the transmit signals. However, for real-world applications, the following two difficulties 
should be addressed  when jointly designing passive and active beamforming: 
\begin{enumerate}
    \item \emph{The joint passive and active beamforming should be designed 
    without cooperating with the illegitimate DIRS;}
    \item  \emph{The joint passive and active beamforming should be designed without 
    the instantaneous CSI of the DIRS-based channels which are unavailable to the AP.}
\end{enumerate}

Given the two practical difficulties, we investigate legitimate IRS-assisted 
anti-jamming precoding against DIRS-based FPJs. The main contributions are summarized as follows:
\begin{itemize}
\item A legitimate IRS is introduced to suppress DIRS-based ACA interference, 
and a practical IRS model is considered in which the reflecting phase shifts of both the DIRS and the legitimate IRS are discrete.  
Moreover, we consider the above two real-world difficulties and then design the passive beamforming at the legitimate IRS by maximizing the received power
of the transmit signals at each LU, which reduces the strength of the DIRS-based ACA
interference relative to that of the transmit signals.
\item To address time-varying DIRS 
phase shifts and the unknown CSI of the DIRS-related channels, we first derive the statistical characteristics of 
the DIRS-jammed channels. Then, a more explicit expression for the signal-to-jamming-plus-noise
ratios (SJNRs) is given. In addition, we also design the active beamforming at the AP
to maximize the SJNRs given the configuration of the legitimate IRS passive beamforming. 
\end{itemize}

The rest of this paper is organized as follows.
In Section~\ref{Princ}, we first describe an MU-MISO system jammed by a temporal DIRS-based FPJ, 
where a legitimate IRS is introduced to mitigate the DIRS-based ACA interference.
We also define the SJNR to quantify performance.
In Section~\ref{ChaMod}, 
the various channel models are defined and  the optimization problem is formulated.
In Section~\ref{StaChara}, we first 
derive a more-explicit form for the SJNR based on the statistical characteristics of the DIRS-jammed channels. 
Then, the active 
beamforming at the AP is designed. In Section~\ref{AntiPre}, the passive beamforming at the legitimate IRS is 
designed via the Riemannian conjugate gradient (RCG) algorithm.
In Section~\ref{ResDis}, simulation results are presented to evaluate the effectiveness of the proposed IRS-enhanced anti-jamming precoder.
Finally, conclusions are given in Section~\ref{Conclus}.

\emph{Notation:} We employ bold capital letters for a matrix, e.g., ${\bf{W}}$, 
lowercase bold letters for a vector, e.g., $\boldsymbol{w}_{k}$, and italic letters for a scalar, e.g., $K$. The superscripts $(\cdot)^{T}$ and $(\cdot)^{H}$ represent the transpose and the Hermitian transpose, respectively, and the symbols $\|\cdot\|$ and $|\cdot|$  represent the Frobenius norm and the absolute value, respectively. 

\section{System Description}\label{Princ}

\subsection{Temporal DIRS-based Fully-Passive Jamming}\label{DIRSFPJ}

\begin{figure}[!t]
    \centering
    \includegraphics[scale=0.58]{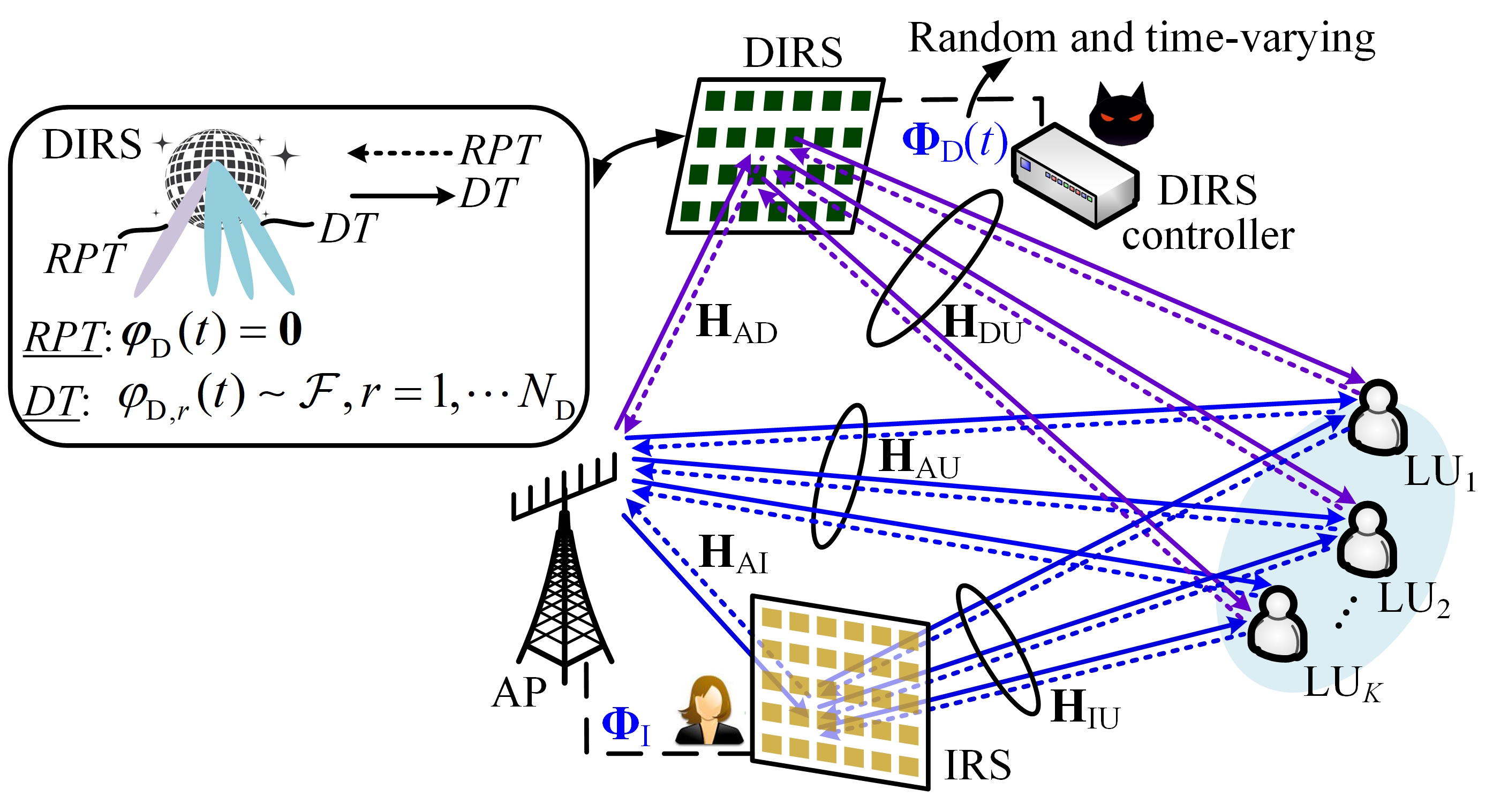}
    \caption{The downlink of an MU-MISO system jammed by a temporal DIRS-based FPJ, 
    where a legitimate IRS is introduced to resist the fully passive attacks launched by the temporal DIRS-based FPJ.}
    \label{fig1}
\end{figure}

Fig.~\ref{fig1} schematically illustrates the downlink of an MU-MISO system in which
a legitimate IRS with $N_{\rm I}$ reflective elements is controlled by
an $N_A$-antenna AP to mitigate  the fully-passive jamming launched by a temporal DIRS-based FPJ
using an $N_{\rm D}$-element illegitimate IRS~\cite{DIRSTWC,MyGC23}.  
We assume that the AP communicates with $K$ single-antenna LUs 
denoted as LU$_1$, LU$_2$, $\cdots$, LU$_K$, and the transmit symbol  
for LU$_k$ ($1\le k \le K$) during the $DT$ phase satisfies 
${\mathbb{E}}\!\left[ \left| s_{{\!D\!T},k} \right|^2 \right] = 1$. 
Consequently, the received signal at LU$_k$ is given by
\begin{alignat}{1}
\nonumber
{y_{\!D\!T,k}}\!\! &= \boldsymbol{h}\!_{D\!T,k}^H\!\sum_{u = 1}^K {{\boldsymbol{w}_{u}}{s_{\!{D\!T,u}}}}  + {n_k}\\
&= \!\!\left(\! { { \boldsymbol{h}_{{\rm{IU}},k}^H  {\!{\bf{\Phi}}_{\rm I}} {{\bf{H}}_{\rm {\!AI}}}  }   \! \!  +\!\!  \boldsymbol{h}_{{\rm{A\!U}},k}^H \!\!   +\!\!   { \boldsymbol{h}_{{\rm{D\! U}},k}^H  {\!{\bf{\Phi}}\!_{\rm D} \!\!\left(t\right)} {{\bf{H}}_{\rm {\!A\!D}}}  } }\!  \right)\!\!\! \sum_{u \! =\!  1}^K {\! {\boldsymbol{w}_{u}}{s_{\!{D\!T,u}}}} \!\!   +\!\!  {n_k} ,
\label{RecSig}
\end{alignat}
where ${\boldsymbol{w}}_k$ is the transmit precoder for user $k$, $n_k \sim CN(0,\sigma^2)$ is additive Gaussian noise,  $\boldsymbol{h}_{{\rm{IU}},k} \! \in \! \mathbb{C}^{ N_{\rm I} \times 1}$ denotes the 
channel between the IRS and LU$_k$, ${{\bf{H}}_{\rm {\!AI}}} \! \in \! \mathbb{C}^{N_{\rm I} \times N_{\rm A}}$
denotes the channel between the AP and the IRS,
$\boldsymbol{h}_{{\rm{A\!U}},k} \! \in \! \mathbb{C}^{N_{\rm A} \times 1}$  denotes 
the channel between the AP and LU$_k$, 
$\boldsymbol{h}_{{\rm{D\!U}},k} \! \in \! \mathbb{C}^{ N_{\rm D} \times 1}$ denotes
the channel between the DIRS and LU$_k$, 
 ${{\bf{H}}_{\rm {\!AD}}} \! \in \! \mathbb{C}^{N_{\rm D} \times N_{\rm A}}$
denotes the channel between the DIRS and the AP, and  
${{\bf{\Phi}}_{\rm I}} = {\rm{diag}}\!\left( {{\boldsymbol{\varphi}}}_{\rm I} \right) 
= {\rm{diag}}\!\left( e^{{{\varphi}}_{{\rm I},1}},\cdots,e^{{{\varphi}}_{{\rm I},{N_{\rm \!I}}}} \right)$
and  
$ {{\bf{\Phi}}_{\rm D} (t) } = {\rm{diag}}\!\left( {{\boldsymbol{\varphi}}}_{\rm D} (t) \right)
= {\rm{diag}}\!\left( e^{{{\varphi}}_{{\rm \!D},1}\!(t)},\cdots,e^{{{\varphi}}_{{\rm \!D},{N_{\rm \!D}}}\!(t)} \right)$ 
represent the IRS reflecting vector and the time-varying DIRS reflecting vector 
during the \emph{DT} phase. For ease of presentation, 
we denote the time-varying DIRS-jammed channel and the IRS-cascaded channel between the AP and LU$_k$ 
by $\boldsymbol{h}_{{\rm{D}},k} (t) =  \boldsymbol{h}_{{\rm{D\! U}},k}^H 
{\bf{\Phi}}_{\rm D} \!\left( t\right) {\bf{H}}\!_{\rm{A\!D}}$
and $\boldsymbol{h}_{{\rm{I}},k} =  \boldsymbol{h}_{{\rm{I\! U}},k}^H 
{\bf{\Phi}}_{\rm I} {\bf{H}}\!_{\rm{A\!I}}$, respectively.  
Moreover, the overall DIRS-LU channel ${{\bf H}}_{\rm{D\!U}}$,
the overal IRS-LU channel ${{\bf H}}_{\rm{IU}}$,
the overall direct channel ${{\bf H}}_{\rm d}$, 
the overall DIRS-jammed channel ${{\bf H}}_{\rm{D}}$,
and the overall IRS-cascaded channel ${{\bf H}}_{\rm{I}}$
are denoted by ${\bf H}_{{\rm{D\!U}}}^H = [\boldsymbol{h}_{{\rm{D\!U}},1}^H,\cdots,\boldsymbol{h}_{{\rm{D\!U}},K}^H  ]$,
 ${\bf H}_{{\rm{IU}}}^H =  [\boldsymbol{h}_{{\rm{IU}},1}^H,\cdots,\boldsymbol{h}_{{\rm{IU}},K}^H  ]$,
  ${\bf H}_{{\rm AU}}^H 
=  [\boldsymbol{h}_{{\rm{AU}},1}^H,\cdots,\boldsymbol{h}_{{\rm{AU}},K}^H  ]$, 
 ${\bf H}_{{\rm{D}}}^H =  [\boldsymbol{h}_{{\rm{D}},1}^H,\cdots,\boldsymbol{h}_{{\rm{D}},K}^H ]$, 
 and ${\bf H}_{{\rm{I}}}^H = [\boldsymbol{h}_{{\rm{I}},1}^H,\cdots,\boldsymbol{h}_{{\rm{I}},K}^H ]$.

An IRS is an ultra-thin surface consisting of massive reflective elements whose 
phase shifts and amplitudes are controlled by simple programmable PIN or varactor diodes~\cite{TCui}.
We will assume the use of PIN diodes, whose ON/OFF behavior only allows for discrete phase shifts.
Furthermore, the DIRS reflecting phase shifts
${\varphi}_{{\rm \!D},r}$ ($r=1,\cdots, N_{\rm \!D}$) are randomly selected from 
a discrete set ${\mathcal{Q}}_{\!D}$ with $b_1$-bit quantized phase shifts 
$\left\{ {\omega_{1}},\cdots,  {\omega_{2^{b_1}}} \right\}$
and follow a stochastic distribution $\cal F$, i.e., ${\varphi}_{{\rm \!D},r} \sim {\cal F}\!\left(\! {\mathcal{Q}}_{\rm D}\right)$.
Meanwhile, the IRS reflecting phase shifts ${\varphi}_{{\rm I},r}$ ($r=1,\cdots, N_{\rm I}$) 
also come from a discrete and finite set ${\mathcal{Q}}_{\!I} = \left\{ {\psi_{1}},\cdots,  {\psi_{2^{b_2}}} \right\}$
with $b_2$-bit quantized phase shifts.

In a channel coherence interval, prior to data transmission, CSI for ${\bf H}_{{\rm{IU}}}$, ${\bf H}_{{\rm{AI}}}$, 
and ${\bf H}_{{\rm{d}}}$  is acquired during the \emph{RPT} phase~\cite{PilotConta,DaipartI}. 
 We assume that the DIRS remains
``silent" during the \emph{RPT} phase~\cite{DIRSTWC,MyGC23}, i.e.,
the DIRS-jammed channel ${\bf H}_{{\rm{D}}}\!(t)$ does not exist during the \emph{RPT} phase. 
As long as the period during which the DIRS phase shifts are changing is about the same as the length of  
the \emph{RPT} phase, the legitimate AP cannot acquire 
any useful knowledge about ${\bf H}_{{\rm{D}}}\!(t)$ via retraining~\cite{MyGC23Extension}. 
Therefore, the temporal DIRS-based FPJ introduces ACA interference during the \emph{DT} phase.

To quantify the impact of the DIRS-based ACA interference, we also define the SJNR for LU$_k$ 
represented by $\eta_k$ with reference to the work in~\cite{MyGC23,MyGC23Extension} and the definition of 
the signal-to-leakage-plus-noise ratio~\cite{RefSLNRadd}, i.e.,
\begin{equation}
    {\eta _k} = \frac{{\mathbb{E}}\!\!\left[{{{\left| {\boldsymbol{h}\!_{D\!T,k}^H{\boldsymbol{w}_{k}}} \right|}^2}}\right]}{{\sum\limits_{u \ne k} \!{\mathbb{E}}\!\!\left[{{{\left| {\boldsymbol{h}\!_{D\!T,u}^H{\boldsymbol{w}_{k}}} \right|}^2} }\right] }+ {\sigma^2} }.
    \label{eqSLNR}
\end{equation}
Following~\cite{RefSLNRadd}, we use a sum-rate-like metric 
based on the SJNR instead of signal-to-interference-plus-noise ratio: 
$R = \sum\nolimits_{k = 1}^K {{{\log }_2}\!\left( {1 + {\eta _k}} \right)}$.
We will refer to this as the SJNR rate.

\subsection{Channel Model and Problem Formulation}\label{ChaMod}
Based on the existing work in~\cite{DIRSVT,DIRSTWC,MyGC23,MyGC23Extension}, 
a DIRS should be positioned near the legitimate AP to 
maximize the jamming impact. 
Moreover, the DIRS must contain a large number of reflective elements to 
cope with the multiplicative large-scale channel fading in the DIRS-jammed channel.
Therefore, the AP-DIRS channel ${\bf H}_{\!{\rm{A\!D}}}$ is constructed based on 
the near-field model~\cite{NearfieldMo1}, i.e., 
\begin{equation}
    \begin{split}
    {\bf H}_{\!{\rm{A\!D}}} \!\!=\!\! {\sqrt{\!\!{\mathscr{L}}_{\rm{\!A\!D}}}} \! 
    \left(\! \sqrt {\frac{{{\varepsilon_{\rm{\!A\!D}}}}}{{1\!+\!{\varepsilon_{\rm{\!A\!D}}}}}}\!{{\widehat{\bf{H}}}^{\!{\rm{LOS}}}}_{\rm{\!A\!D}}  \!+\! 
    \sqrt {\frac{1}{{1\!+\!{\varepsilon_{\rm{\!A\!D}}}}}}\!{{\widehat {\bf{H}}}^{{\rm{NLOS}}}_{\rm{\!A\!D}}} \! \right),
    \end{split}
\label{Ricianchan}
\end{equation}
where ${\mathscr{L}}_{\rm{\!A\!D}}$ denotes the large-scale channel fading between 
the AP and the DIRS, and ${{\varepsilon_{\rm{\!A\!D}}}}$ represents the Rician factor for ${\bf H}_{\!{\rm{A\!D}}}$,  
and ${{\widehat {\bf{H}}}_{\rm{\!A\!D}}^{{\rm{NLOS}}}}$ is assumed to follow Rayleigh fading~\cite{QQWu,BookFarFeild}, i.e.,
$\left[ {{\widehat {\bf{H}}}_{\rm{\!A\!D}}^{{\rm{NLOS}}}} \right]_{r,n} \sim \mathcal{CN}\left(0,1\right), r=1, \cdots, N_{\rm D}$ and $n=1, \cdots, N_{\rm A}$.
The elements of 
 the line-of-sight (LOS) channel ${{\widehat {\bf{H}}}_{\rm{\!A\!D}}^{{\rm{NLOS}}}}$ 
 are given by~\cite{MyGC23,MyGC23Extension,NearfieldMo1}
 \begin{equation}
    \left[{{\widehat {\bf{H}}}_{\rm{\!A\!D}}^{{\rm{NLOS}}}}\right]_{r,n} = {e^{ - j\frac{{2\pi }}{\lambda }\left( {{D_n^r} - {D_n}} \right)}},
    \label{GLOS}
\end{equation}
where $\lambda$ is the wavelength of the transmit signals, 
and ${D_n^r}$ and ${D_n}$ represent the distance between 
the $n$-th antenna and the $r$-th DIRS reflective element 
and the distance between the $n$-th antenna and the centre (origin) of the DIRS, respectively. 
We will assume that the distances between two adjacent transmit antennas and 
two adjacent DIRS reflective elements are both $d ={\lambda}/2$. 
    
The DIRS-LU channel ${\bf H}_{\!{\rm{D\!U}}}$ and the direct AP-LU channel ${\bf H}_{\!{\rm{A\!U}}}$ 
are modeled based on Rayleigh fading:
\begin{alignat}{1}
    &{{\bf{H}}_{\rm{\!D\!U}}} \!=\! {{\bf{L}}_{\rm{\!D\!U}}^{ \frac{1}{2}}}{\widehat {\bf{H}}_{\rm{\!D\!U}}} 
     \!=\!\! \left[\! {{\sqrt{\!{\!{\mathscr{L}}_{{\rm{D\!U}},1}}}}{{\widehat {\boldsymbol{h}}}_{{\rm{D\!U}},1}},  
    \!\cdots , \!{\sqrt{\!{\!{\mathscr{L}}_{{\rm{D\!U}},K}}}}{{\widehat {\boldsymbol{h}}}_{{\rm{D\!U}},K}}} \right], \label{HIkeq}\\
    &{{\bf{H}}_{\rm{\!A\!U}}} \!=\! {{\bf{L}}_{\rm{\!A\!U}}^{ \frac{1}{2}}}{\widehat {\bf{H}}_{\rm{\!A\!U}}} 
    \!=\!\! \left[\! {{\sqrt{\!{\!{\mathscr{L}}_{{\rm{A\!U}},1}}}}{{\widehat {\boldsymbol{h}}}_{{\rm{A\!U}},1}}, 
    \! \cdots, \! {\sqrt{\!{\!{\mathscr{L}}_{{\rm{A\!U}},K}}}}{{\widehat {\boldsymbol{h}}}_{{\rm{A\!U}},K}}} \right],
    \label{Hdkeq}
\end{alignat}
where the elements of 
${\bf{L}}_{\rm{D\!U}} \!=\! {\rm{diag}}\!\left({{\mathscr{L}}_{{\rm{D\!U}},1}},\cdots,{{\mathscr{L}}_{{\rm{D\!U}},K}}\right)$ and ${\bf{L}}_{\rm{A\!U}} = {\rm{diag}}\!\left({{\mathscr{L}}_{{\rm{A\!U}},1}},\cdots,{{\mathscr{L}}_{{\rm{A\!U}},K}}\right)$ 
denote the large-scale channel fading coefficients, which are assumed to be independent. 
The elements of ${\widehat {\bf{H}}_{\rm{D\!U}}}$ and ${\widehat {\bf{H}}_{\rm{A\!U}}}$ 
are assumed to be independent and identically distributed (i.i.d.) Gaussian random variables~\cite{BookFarFeild}, i.e., 
$\left[{\widehat {\bf{H}}_{\rm{D\!U}}}\right]_{r,k},\left[{\widehat {\bf{H}}_{\rm{A\!U}}}\right]_{n,k} 
\sim \mathcal{CN}\left(0,1\right), r=1, \cdots, N_{\rm D}$, $n=1, \cdots, N\!_{\rm A}$, and $k = 1, \cdots,K$.

The legitimate IRS is carefully pre-positioned, 
ensuring that the cascaded IRS-aided AP-LU links are not blocked~\cite{QQWu,MyEE}. 
Therefore, the AP-IRS channel ${\bf H}_{\!{\rm{A I}}}$ and the IRS-LU channel ${\bf H}_{ {\rm{I U}}}$
are assumed to follow Rician fading, which is modeled as~\cite{DIRSVT,QQWu,MyEE}
\begin{alignat}{1}
&{{\bf{H}}_{\rm{A\!I}}}\! = \!{\sqrt{\!{\mathscr{L}}_{\rm{A\!I}}}} \!\left(\!\!{\sqrt {\frac{{{\varepsilon_{\rm{\!A\!I}}}}}{{1\!+\!{\varepsilon_{\rm{\!A\!I}}}}}} {{\widehat{\bf{H}}}^{\!{\rm{LOS}}}}_{\rm{\!A\!I}}\!+\!\sqrt {\frac{1}{{1 + {\varepsilon_{\rm{\!A\!I}}}}}} {{\widehat{\bf{H}}}^{\!{\rm{NLOS}}}}_{\rm{\!A\!I}}}\!\!\right),\label{HAIoa}\\
&{{\bf{H}}_{\rm{I\!U}}}  \!=\! \left[\! {{\sqrt{{{\mathscr{L}}_{{\rm{I\!U}},1}}}}{{\widehat {\boldsymbol{h}}}_{{\rm{I\!U}},1}},  \cdots ,{\sqrt{{{\mathscr{L}}_{{\rm{I\!U}},K}}}}{{\widehat {\boldsymbol{h}}}_{{\rm{I\!U}},K}}} \right],\label{HIUk}
\end{alignat}
where ${\mathscr{L}}_{\rm{A\!I}}$ and ${\mathscr{L}}_{{\rm{I\!U}},k}$ respectively represent the large-scale channel fading coefficients
between the AP and the IRS and between the IRS and LU$_k$,  and ${\varepsilon_{{\rm{A\!I}}}}$ is 
the Rician factor for ${{\bf{H}}_{\rm{A\!I}}}$. 
Moreover, ${{\widehat {\boldsymbol{h}}}_{{\rm{I\!U}},k}}$ can be expressed by 
\begin{equation}
    {{\widehat {\boldsymbol{h}}}_{{\rm{I\!U}},k}} = \!\left(\!\!{\sqrt {\frac{{{\varepsilon_{{\rm{I\!U}},k}}}}{{1\!+\!{\varepsilon_{{\rm{I\!U}},k}}}}} 
    {{\widehat{\boldsymbol{h}}}^{\!{\rm{LOS}}}}_{{\rm{I\!U}},k}\!+\!\sqrt {\frac{1}{{1 \!+\! {\varepsilon_{\rm{I\!U}}}}}} 
    {{\widehat{\boldsymbol{h}}}^{\!{\rm{NLOS}}}}_{{\rm{I\!U}},k} }\!\right).
     \label{hIUk}
\end{equation}
where ${\varepsilon_{{\rm{I\!U}},k}}$ denotes the Rician factor of the channel ${\boldsymbol{h}}_{{\rm{I\!U}},k}, k =1,\cdots, K.$  
The LOS channels ${{\widehat{\bf{H}}}^{\!{\rm{LOS}}}}_{\rm{\!A\!I}}$ and ${{\widehat{\boldsymbol{h}}}^{\!{\rm{LOS}}}}_{{\rm{I\!U}},k}$
are written as~\cite{DIRSVT,QQWu}
\begin{alignat}{1}
    &{{\widehat{\bf{H}}}^{\!{\rm{LOS}}}}_{\rm{\!A\!I}} = \!\sqrt {\!{N_{\rm{\!I}}}{N_{\rm\!A}}} {\boldsymbol{\alpha} _{\rm{I}}}\!\left( {{\vartheta},{\theta}} \right)\boldsymbol{\alpha} _{\rm{\!A}}^H\!\!\left( {{\phi}} \right),
    \label{LOS1}\\
    &{{\widehat{\boldsymbol{h}}}^{\!{\rm{LOS}}}}_{{\rm{I\!U}},k} = 
    \!\sqrt {\!{N_{\rm{I}}}} {\boldsymbol{\alpha} _{\rm{I}}}\!\left( {{\vartheta _{ k}},{\theta_{ k}}} \right),
    \label{LOS2}
\end{alignat}
where $\boldsymbol{\alpha} _{\rm{\!A}}$ and $\boldsymbol{\alpha} _{\rm{I}}$ represent the array responses.

In this work, we aim to maximize the performance metric $R$ obtained from the MU-MISO system
by optimizing the transmit precoder ${\bf W} = \left[{\boldsymbol{w}}_{1},\cdots, {\boldsymbol{w}}_{k} \right]$ 
and the IRS reflecting vector ${\boldsymbol{\varphi}}_{\rm I}$. 
For a practical IRS, its gains are the tion of its corresponding phase shifts~\cite{IRSPSGV}. 
Mathematically, the optimization problem is expressed as 
\begin{alignat}{1}
    \left( {{\rm{P}}1} \right): &\mathop {\max}\limits_{{\boldsymbol{\varphi}_{\rm I}},{\bf{W}}} R = \sum\limits_{k = 1}^K {{{\log }_2}\!\left( {1 + {\eta _k}} \right)}\\
    &{\rm{s}}.{\rm{t}}.\; {\left\| {\bf{W}} \right\|^2} \le P_0,
    \label{addeq4}\\
    &\;\;\;\;\;\; {\boldsymbol{\varphi}_{\rm I}} =  {\rm{diag}}\!\left( e^{{\varphi}_{{\rm I},1}}, \cdots, e^{{{\varphi}_{{\rm I},{N_{\rm \!I}}}}} \right),
    \forall {\varphi}_{{\rm I},r} \in {\mathcal{Q}}_{\!I}.
    \label{OP1}
\end{alignat}

$({\rm{P}}1)$ is a typical mixed integer nonlinear programming (MINLP) problem, which is NP-hard. 
In addition, the DIRS-jammed channel ${\bf H}_{\rm {\!D}}$ includes the time-varying i.i.d. random variables $\left\{{{\varphi}_{{\rm D},1}\!(t)}, \cdots,{{\varphi}_{{\rm \!D},{N_{\rm \!D}}}\!(t)}\right\}$ 
in each channel coherence interval. Moreover, the legitimate AP cannot train to learn the CSI of the DIRS-related channels ${\bf H}_{\rm {\!A\!D}}$ and ${\bf H}_{\rm {\!D\!U}}$. 
These difficulties make $({\rm{P}}1)$ very challening to solve. 
We assume that the perfect CSI of 
the AP-IRS channel ${\bf H}_{\rm {\!AI}}$, the IRS-LU channel ${\bf H}_{\rm {I\!U}}$, and the direct AP-LU channel ${\bf H}_{\rm {A\!U}}$ 
is available via training with the IRS and the LUs during the \emph{RPT} phase~\cite{PerfectIRSCSI}, 
as imperfect CSI is not a primary concern in the jamming scenario here, 
and its impact has also been thoroughly studied~\cite{RefSLNRadd,LDetector1}.

\section{IRS-Enhanced Anti-Jamming Precoding Against DIRS-Based FPJ}\label{EnhaAPJ}

\subsection{Active Beamforming Design}\label{StaChara}
Based on the definition of $\eta_k$ in~\eqref{eqSLNR}, the time-varying reflecting vector 
${\boldsymbol{\varphi}_{\rm \!D}}\!(t)$ and the lack of CSI of the DIRS-related channels 
${\bf H}_{\rm {\!A\!D}}$ and ${\bf H}_{\rm {\!D\!U}}$, it is infeasible to solve $({\rm{P}}1)$ directly.
To find a more tractable solution,  we derive a more-explicit expression for $\eta_k$.
\newtheorem{proposition}{Proposition}
\begin{proposition}
    \label{Proposition1}
    Under the condition that ${\boldsymbol{h}}_{{\rm{\!I}},k}$,  ${\boldsymbol{h}}_{{\rm{\!D}},k}\!(t)$, 
    and ${\boldsymbol{h}}_{{\rm{\!d}},k}$
    are independent and ${\boldsymbol{h}}_{{\rm{\!I}},k}$ and ${\boldsymbol{h}}_{{\rm{\!d}},k}$ 
    are unchanged during channel coherence time, 
    $\eta_k$ defined in~\eqref{eqSLNR} reduces to
    \begin{alignat}{1}
        \nonumber
        &{{\overline \eta  _k}} \!\! = \! \frac{ \! {{{  {\boldsymbol{w}_{k}^H}\!\! \left(\! 
        \left(  \!\boldsymbol{h}_{{\rm{\!I}},k} \!\! + \! \boldsymbol{h}_{{\rm{\!d}},k} \!\right) \!
        \left(\!\boldsymbol{h}_{{\rm{\!I}},k} \!\! + \! \boldsymbol{h}_{{\rm{\!d}},k} \!\right)^{\!H}  
        \!\!\!+\! {\beta _k}{{\bf I}\!_{ N\!_ {\rm A}}} \!\right)\!\!{\boldsymbol{w}\!_{k}} } }} } 
        {\!{\sum\limits_{u \ne k} 
        \!\! \! {{{  {\boldsymbol{w}_{k}^H}\!\! \left(\!  
        \left(\!  \boldsymbol{h}_{{\rm{\!I}},u} \!\! + \! \boldsymbol{h}_{{\rm{\!d}},u} \!\right) 
        \!\left(\!\boldsymbol{h}_{{\rm{\!I}},u} \!\! + \! \boldsymbol{h}_{{\rm{\!d}},u}\! \right)^{\!H} 
        \!\!\!+\! {\beta _k}{{\bf I}\!_{ N\!_ {\rm A}}} \!\right)\!\!{\boldsymbol{w}\!_{k}}   } }} } \! +\! {\sigma^2} }, \\
        &\;\;\;\;\;\;\;\;\;\;\;\;\;\;\;\;\;\;\;\;\;\;\;\;\;\;\;\;\;\;\;\;\; 
        \;\;\;\;\;\;\;\;\;\;\;\;\;\;\;\;\;\;\;\;\;\;\;\; {\rm{as}}\;N_{\rm \!D} \to \infty,
        \label{eqSLNRredadd}
    \end{alignat}
    where ${\beta _k} = {{{\mathscr{L}}\!_{{\rm G}}}{{\mathscr{L}}\!_{{\rm I},k}}{N\!_{\rm D}} }$, and ${\bf I}_{\!{N_{\rm \!A}}}$ is an $N_{\rm \!A} \!\times\! N_{\rm \!A}$ unit matrix.
\end{proposition}

\begin{IEEEproof}
    See Appendix~\ref{AppendixA}.
\end{IEEEproof}

Consequently,  $({\rm{P}}1)$ reduces to the following form:
\begin{alignat}{1}
    \left( {{\rm{P}}2} \right): &\mathop {\max}\limits_{{\boldsymbol{\varphi}_{\rm I}},{\bf{W}}} {\overline R}  
    = \sum\limits_{k = 1}^K {{{\log }_2}\!\left( {1 + {{\overline \eta  _k}}} \right)}\\
    \nonumber
    &{\rm{s}}.{\rm{t}}.~\eqref{addeq4}~\mbox{and}~\eqref{OP1}.
\end{alignat}
When the IRS reflecting vector ${\boldsymbol{\varphi}_{\rm I}}$ is determined, $({\rm{P}}2)$ reduces to the following subproblem:
\begin{alignat}{1}
\left( {{\rm{P}}2-{\rm S}1} \right): &\mathop {\max}\limits_{{\bf{W}}} {\left. {{\overline R} } \right|_{{\boldsymbol{\varphi}_{\rm I}}}} \!= \!\! \sum\limits_{k = 1}^K {{{\log }_2}\!\left( \!{1 + {\left. {\overline \eta  _k}\right|_{{\boldsymbol{\varphi}_{\rm I}}} }} \!\right)}\\
\nonumber
&{\rm{s}}.{\rm{t}}.~\eqref{addeq4}.
\end{alignat}
Based on the work in~\cite{MyGC23,MyGC23Extension}, an anti-jamming precoder 
 ${\bf W}^{\rm{o}}$ for $\left( {{\rm{P}}2-{\rm S}1} \right)$
which can maximize the SJNRs is given by
\begin{equation}
    {\boldsymbol{w}{{_{k}^{\rm o}}}} \propto \max.{\rm{eigenvector}}\left({\bf A}_k\right),
    \label{AntiJamm}
\end{equation}
where
\begin{equation}
    {\bf A}_{\!k} \!\!= \!\!\!\left(\!{{\boldsymbol{h}_{{\rm \!L},k}}\boldsymbol{h}_{{\rm \!L},k}^H \!+\! {{\widetilde \beta _k}}{{\bf{I}}\!_{N\!_{\rm \!A}}}}\!\right)  
    \!\!\!\left(\!\!{{ {{{\widetilde {\bf{H}}}_{{\rm \!L},k}}\widetilde {\bf{H}}_{{\rm \!L},k}^H \!\!+\! \!\!\left( \!\!{\frac{{{\sigma ^2}}\!K}{{{P_0}}} \!\!+ \!\!\! {\sum\limits_{u \ne k}\!{{\widetilde \beta _u}}  }}\! \!\right)\!\!{{\bf{I}}\!_{N\!_{\rm \!A}}}}  }}\!\!\right)^{\!\!{\!-1}}\!\!,
    \label{AntiJammMat}
\end{equation}
${{\widetilde {\bf{H}}}_{{\rm \!L},k}} \!=\! \left[{\boldsymbol{h}_{{\rm \!L},1}},\cdots,{\boldsymbol{h}_{{\rm \!L},k-1}},{\boldsymbol{h}_{{\rm \!L},k+1}},\cdots,{\boldsymbol{h}_{{\rm \!L},K}} \right]$,  
${\boldsymbol{h}_{{\rm \!L},k}} \!=\! {\boldsymbol{h}_{{\rm \!I},k}} + {\boldsymbol{h}_{{\rm \!d},k}}$, and 
$\{{{\widetilde \beta _k}} \}_{k = 1}^K$ are estimates of $\left\{{\beta _k}\right\}_{k=1}^K$ in~\eqref{eqSLNRredadd}.
A  frame structure has been designed in~\cite{MyGC23Extension}  to acquire the estimates
since it is unrealistic for the AP to know the values of $\left\{{\beta _k}\right\}_{k=1}^K$ in advance. 
Therefore, only the estimated $\{{{\widetilde \beta _k}} \}_{k = 1}^K$ can be used 
in the design of the IRS-enhanced anti-jamming precoder.

\subsection{Passive Beamforming Design}\label{AntiPre} 
The work in~\cite{MyWCMag} showed that the legitimate AP can increase the strength of
the transmit signals to counter the DIRS-based ACA attacks.
Inspired by this investigation, we compute ${\boldsymbol{\varphi}_{\rm I}^{\rm o}}$ by maximising the
overall effective channel power $P_{\rm E}$, where the overall effective channel is  ${{\bf H}_{\rm{E}}} \!=\! {{\bf H}_{\rm{I}}} \!+\! {{\bf H}_{\rm{A\!U}}}$, i.e., 
\begin{alignat}{1}
    \left({{\rm{P}}3} \right): &  \mathop {\max}\limits_{{\boldsymbol{\varphi}_{\rm I}}} 
    P_{\rm E} \!=\! {\left\| {{{\bf{H}}_{\rm{E}}}} \right\|^2} \label{R1Objfun}\\
    &{\rm{s}}.{\rm{t}}.~\eqref{OP1}.
    \label{P1R1}
\end{alignat}
However, $({{\rm{P}}3})$ is still difficult to solve due to the discrete variables $\left\{ {{\varphi _{{\rm I},r}}} \right\}_{r = 1}^{{N_{\rm{I}}}}$. 
Therefore, we further relax  $\left\{ {{\varphi _{{\rm I},r}}} \right\}_{r = 1}^{{N_{\rm{I}}}}$ 
to continuous variables. 
As a result, $({{\rm{P}}3})$ is relaxed to 
\begin{alignat}{1}
    \left( {{\rm{P}}3\!-\!{\rm{R}}1} \!\right)\!\!: &\mathop {\max}\limits_{{\boldsymbol{\varphi}_{\rm I}}} 
    P_{\rm \!E} \!=\!\! {\left\| {{{\bf{H}}_{\rm{E}}}} \right\|^2} \label{R1Objfun1}\\
    &{\rm{s}}.{\rm{t}}. {\boldsymbol{\varphi}_{\rm I}} \!=\!  {\rm{diag}}\!\left(\! e^{{\varphi}_{{\rm I},1}},
    \! \cdots,\! e^{{{\varphi}_{{\rm I},{N_{\rm \!I}}}}}\! \right),
    \forall {\varphi}_{{\rm I},r} \!\!\in\! [0,2\pi].
    \label{R1S2}
\end{alignat}


The objective function in~\eqref{R1Objfun1} is a continuous and differentiable function of
the relaxed continuous ${\boldsymbol{\varphi}_{\rm I}}$. Furthermore, the constraint in~\eqref{R1S2}
creates a complex circle manifold. Therefore, we can use 
the RCG algorithm to solve $({{\rm{P}}3\!-\!{\rm{R}}1})$~\cite{DIRSTWC,AORCG}.

\setlength{\topmargin}{-0.721 in}
\begin{algorithm}[t]
    \renewcommand{\algorithmicrequire}{\textbf{Input:}}
    \renewcommand{\algorithmicensure}{\textbf{Initialize:}}
    \caption{IRS-enhanced anti-jamming precoding algorithm} 
    \label{alg:1}
    \begin{algorithmic}[1]
      \REQUIRE ${\bf{H}}_{\rm{\!I\!U}}$, ${\bf{H}}_{\rm{\!A\!I}}$, ${\bf{H}}_{\rm{\!A\!U}}$,
      $\{{{\widetilde \beta _k}} \}_{k = 1}^K$, $K$, $N\!_{\rm \!A}$, and $P_0$. 
      \ENSURE ${\bf{\Phi}}_{\rm I}  = {\bf{I}}_{\! N_{\rm I}}$. 
      \STATE  Compute ${\boldsymbol{\varphi}_{\rm I}^{R\!C\!G}}$ by solving $({{\rm{P}}3-{\rm{R}}1})$;
      \STATE  Calculate ${\boldsymbol{\varphi}_{\rm I}^{\rm o}}$ by solving $({{\rm{P}}4 })$;
      \STATE  Substitute ${\boldsymbol{\varphi}_{\rm I}^{\rm o}}$ to $({{\rm{P}}2-{\rm{S}}1})$ 
      \STATE  Compute ${\bf W}^{\rm o}$ by~\eqref{AntiJamm};
      \renewcommand{\algorithmicrequire}{\textbf{Output:}}
      \REQUIRE  {${\bf W}^{\rm o}  $ and ${\boldsymbol{\varphi}_{\rm I}^{\rm o}}  $}.
    \end{algorithmic}
\end{algorithm}

1) Riemannian Gradient: The Riemannian gradient ${\rm{grad}}\!\left( {P_{\rm E}} \right) $ at the relaxed ${\boldsymbol{\varphi}_{\rm I}}$  
is a tangent vector that denotes the greatest decreasing direction 
of $P_{\rm E}$, i.e.,
\begin{equation}
    {\rm{grad}}\!\left( P_{\rm E}\right) 
    \!= \!\nabla\!\!\left(  { P_{\rm E} }  \right) 
    \!-\! {\mathop{\rm Re}\nolimits} \!\left\{ \!{\nabla\!\! \left(  { P_{\rm E} }  \right) 
     \!\odot\!{{\boldsymbol{\varphi}}_{\rm \!I}^H}} \!\right\} \! \odot\! {{\boldsymbol{\varphi}}_{\rm \!I}},
    \label{RCG1}
\end{equation}
where $\nabla\!\! \left(P_{\rm E}\right)$
represents the Euclidean gradient.

2)  Search Direction: 
The tangent vector conjugate to ${\rm{grad}}\!\left( P_{\rm E} \right)$ 
can be used as the search direction $\mathcal{D}$, i.e.,
\begin{equation}
\mathcal{D}= -{\rm{grad}}\!\left(P_{\rm E}\right) +{\rho_1 (\tilde{\mathcal{D}}-{\mathop{\rm Re}\nolimits} \{ \!{\tilde{\mathcal{D}}\odot{{\boldsymbol{\varphi}}_{\rm \!I}^H}} \!\} \!\odot\! {\boldsymbol{\varphi}}_{\rm \!I} ) } ,
\label{RCG2}
\end{equation}
where $\rho_1$ and $\tilde{\mathcal{D}}$ denote the conjugate gradient update parameter 
and the previous search direction, respectively.

3) Retraction: The tangent vector is retracted back to 
the complex circle manifold described by~\eqref{R1S2}, i.e.,
\begin{equation}
\frac{{{{\left( {{\boldsymbol{\varphi}}_{\rm \!I} + {\rho _2}\mathcal{D}} \right)}_r}}}{{\left| {{{\left( {{\boldsymbol{\varphi}}_{\rm \!I}  + {\rho _2}\mathcal{D}} \right)}_r}} \right|}} \mapsto {{ {\varphi}}_{{\rm I},r}},
\label{RCG3}
\end{equation}
where $\rho_2$ denotes the Armijo step size and $r =1,\cdots, N_{\rm \!I}$.

Based on the continuous ${\boldsymbol{\varphi}_{\rm I}^{R\!C\!G}}$ obtained from $({{\rm{P}}3-{\rm{R}}1})$,
the discrete solution ${\boldsymbol{\varphi}}^{\rm o}_{\rm \!I}$ is found by choosing the closest value in the discrete phase set:
\begin{alignat}{1}
    \left( {{\rm{P}}4} \right):\; & {\boldsymbol{\varphi}}_{\rm{I}}^{\rm o} 
    = \mathop {\arg \min }\limits_{{{\boldsymbol{\varphi}}_{\rm{I}}}} {\left\| {{{\boldsymbol{\varphi}}_{\rm{I}}} 
    - {\boldsymbol{\varphi}}_{\rm{I}}^{R\!C\!G}} \right\|^2} \label{DescVar}\\
    \nonumber
    &{\rm{s}}.{\rm{t}}.~\eqref{OP1}.
\end{alignat}
Based on the computed ${\boldsymbol{\varphi}}^{\rm o}_{\rm \!I}$, we then calculate  ${\bf W}^{\rm o}$ by~\eqref{AntiJamm}.
The pseudo-code for the proposed IRS-enhanced anti-jamming precoding algorithm is given in Algorithm~\ref{alg:1}. 

\section{Simulation Results and Discussion}\label{ResDis}
In this section, numerical results are presented to evaluate the effectiveness of the IRS-enhanced anti-jamming precoder. 
We consider an MU-MISO system in which the legitimate AP equipped with 32 antennas is assisted by a $(16\! \times \!8)$-element IRS to communicate with 16 single-antenna LUs,
 and the DIRS has $(64\times 32)$  elements,
 i.e., $N_{\rm A} =32$, $N_{\rm I} = 128$, $K=16$, and $N_{\rm D} = 2048$. 
 Similar to~\cite{DIRSVT,DIRSTWC,MyGC23,MyGC23Extension}, we further assume that a one-bit DIRS is implemented, 
 where ${\varphi}_{{\rm D},r} \in {{\cal Q}}_{\rm D} = \left\{ {\frac{\pi}{9},\frac{6\pi}{5}} \right\}$. 
Meanwhile, the IRS phase shifts are quantified by 2 bits, i.e.,
 ${\varphi}_{{\rm I},r} \in {{\cal Q}}_{\rm I} = \left\{ {0, \frac{\pi}{2}, {\pi}, \frac{3\pi}{2}} \right\}$.

The centers of the legitimate AP, the legitimate IRS, and the DIRS are located at (2m, 0m, 5m), (10m, 280m, 5m), and (2m, 0m, 2m), respectively. 
The LUs are randomly distributed in a circular region centred at (0m, 300m, 0m) with a radius of 20m. 
According to the 3GPP propagation model~\cite{3GPP}, the propagation parameters of the wireless channels modelled 
in~Section~\ref{ChaMod} are described as follows:
 ${\mathscr{L}}_{{\rm {A\!D}}},{\mathscr{L}}_{{\rm {A\!I}}}, {\mathscr{L}}_{{\rm {I\!U}},k} \!=\!35.6\!+\!22{\log _{10}}({d_{i}})$ 
 and $ {\mathscr{L}}_{{\rm {A\!U}},k}, {\mathscr{L}}_{{\rm {D\!U}},k} = {\mathscr{L}}_{{\rm I},k}\!=\!32.6\!+\!36.7{\log _{10}}({d_{{i}}})$,
where $d_{i} \in \left\{d_{{\rm {A\!D}}}, d_{{\rm {A\!I}}}, d_{{\rm {I\!U}},k}, d_{{\rm {A\!U}},k},d_{{\rm {D\!U}},k}\right\}$ is the propagation distance. 
Moreover,  the AWGN variance is $\sigma^2\!=\!-170\!+\!10\log _{10}\left(BW\right)$ dBm, and the bandwidth is $BW=180$ kHz. 
If not otherwise specified, the above parameters default to these values.

Herein, we show the ergodic rate per LU (i.e., ${\frac{{R} }{K}}$) achieved by
the following benchmarks: 1) the rate resulting from a traditional MU-MISO system without jamming attacks (W/O Jamming),
2) the rate resulting from an MU-MISO system using the IRS-enhanced anti-jamming precoder presented in Algorithm~\ref{alg:1} (Proposed),
3) the rate resulting from an MU-MISO system using the anti-jamming precoder given in~\cite{MyGC23,MyGC23Extension} (AJP in~\cite{MyGC23,MyGC23Extension}),
4) the rate resulting from an MU-MISO system with a temporal DIRS-based FPJ~\cite{DIRSVT,DIRSTWC} (FPJ in~\cite{DIRSVT,DIRSTWC}) without any anti-jamming, 
and 5) the rate resulting from an MU-MISO system with an AJ located at (2m, 0m, 5m) 
broadcasting AWGN signals with 0 dBm jamming power (W/ $P_{\rm J} =  0$ dBm). 

\begin{figure}[!t]
    \centering
    \includegraphics[scale=0.28]{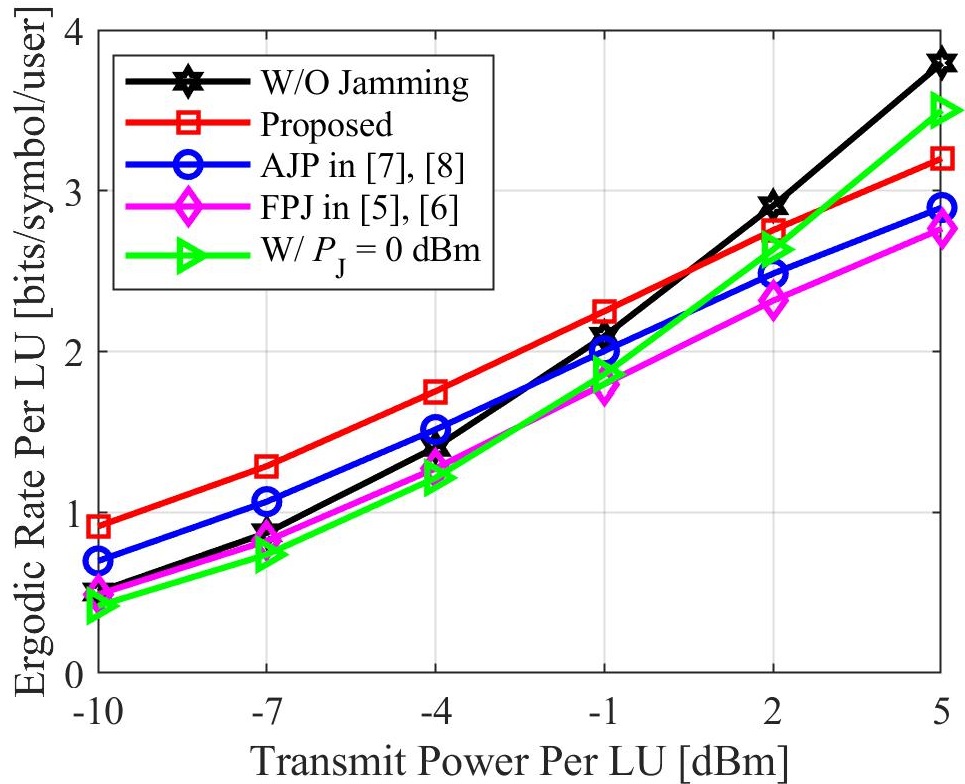}
    \caption{Ergodic rate per LU vs transmit power per LU.}
    \label{fig2}
\end{figure}

Fig.~\ref{fig2} illustrates the ergodic rate as a function of the transmit power per LU. We can see that 
the proposed IRS-enhancedanti-jamming precoder achieves better ergodic rates per LU than the anti-jamming precoder designed in~\cite{MyGC23,MyGC23Extension}.
The work in~\cite{MyGC23,MyGC23Extension} has shown that the anti-jamming precoder can to some extent use 
the DIRS-jammed channels to increase the the ergodic rates of LUs. As a result,  when the transmit power per LU (i.e., $\frac{P_0}{K}$) is low,
the rate from the anti-jamming precoder in~\cite{MyGC23,MyGC23Extension} is higher than that without any jamming.
The precoder proposed in this paper also achieves some gain from the DIRS-jammed channels, 
and it improves the rate by exploiting the legitimate IRS.
Therefore, the proposed IRS-enhanced anti-jamming precoder achieves better performance
compared to the work in~\cite{MyGC23,MyGC23Extension}.

\begin{figure}[!t]
    \centering
 \begin{minipage}{0.49\linewidth}
     \centerline{\includegraphics[width=1.05\linewidth]{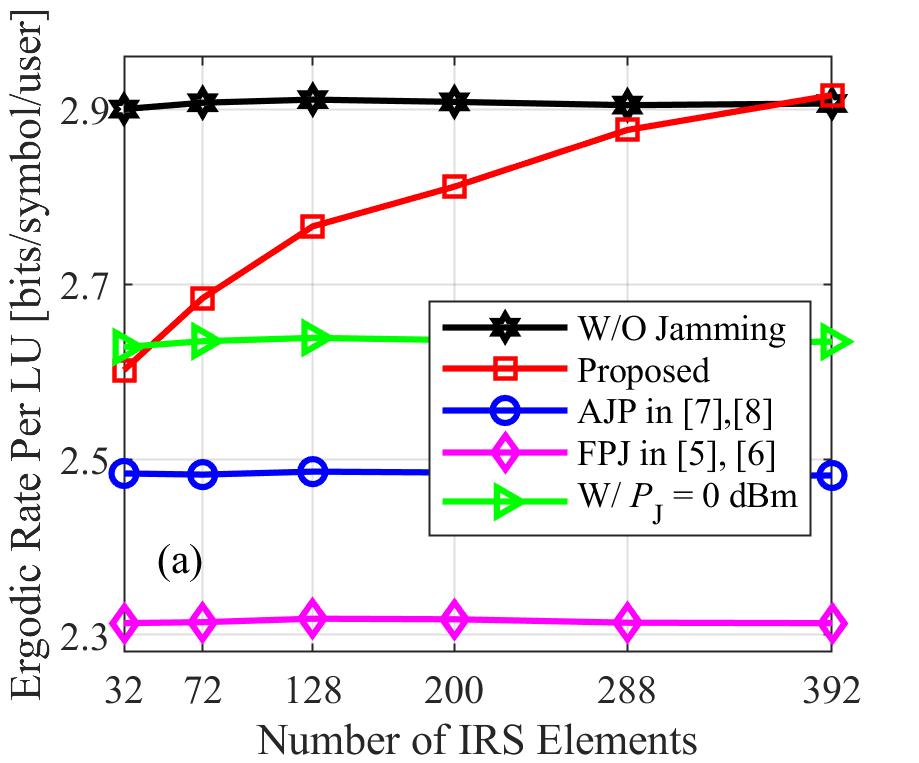}}
     \label{ResFig2la}
 \end{minipage}
    \begin{minipage}{0.49\linewidth}
     \centerline{\includegraphics[width=1.05\linewidth]{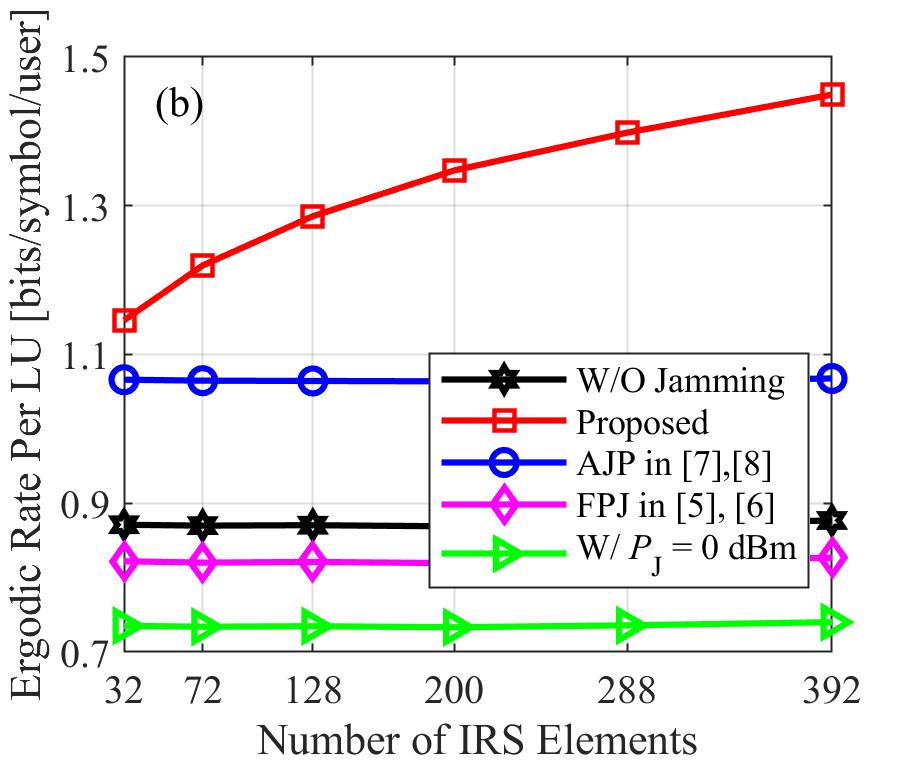}}
     \label{ResFig2lb}
 \end{minipage}
\caption{Relationship between the ergodic rate per LU and the number of IRS reflective elements for different benchmarks, where the transmit power per LU is (a) 0 dBm and (b) -7 dBm, respectively.}
	\label{ResFig2l}
\end{figure}

Fig.~\ref{ResFig2l} (a) and (b) illustrate the influence of the size of the legitimate IRS for high (0 dBm) and low (-7 dBm) transmit power cases, respectively.
As the number of IRS reflective elements increases, the impact of the DIRS-based ACA interference can gradually be diminished out even for high transmit power.
However, the slope of the ergodic rate curve over the proposed precoder decreases as the number of IRS reflective elements continues to increase.
We see that the proposed IRS-enhanced anti-jamming precoder provides a possible approach to mitigate the DIRS-based ACA interference.

\begin{figure}[!t]
    \centering
 \begin{minipage}{0.49\linewidth}
     \centerline{\includegraphics[width=1.05\linewidth]{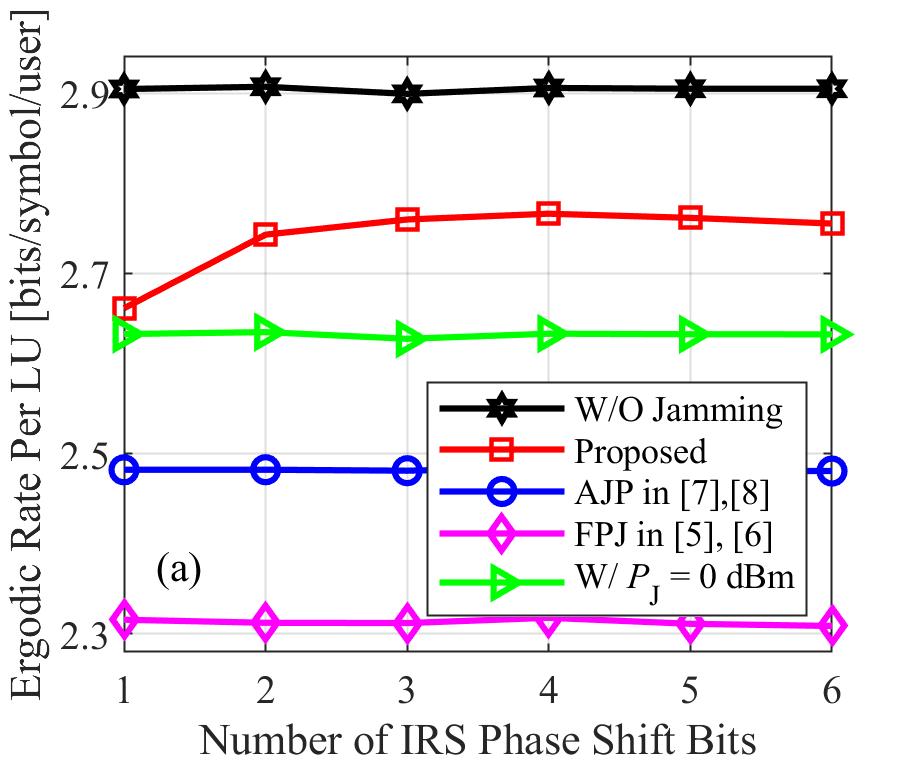}}
     \label{ResFig3la}
 \end{minipage}
    \begin{minipage}{0.49\linewidth}
     \centerline{\includegraphics[width=1.05\linewidth]{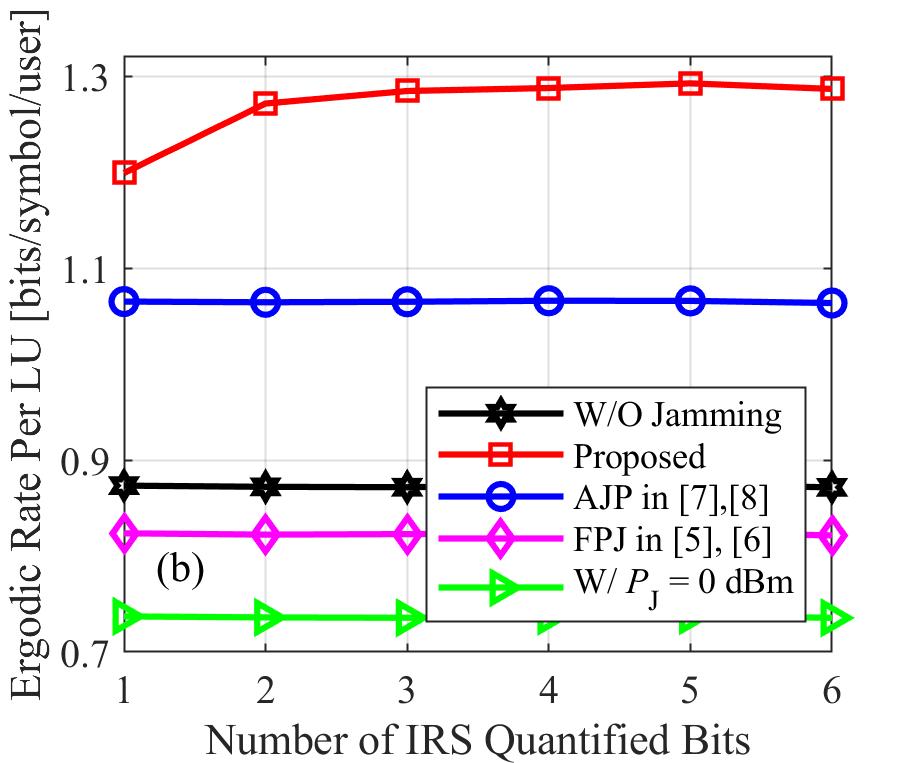}}
     \label{ResFig3lb}
 \end{minipage}
\caption{Relationship between the ergodic rate per LU and the number of IRS quantified bits for different benchmarks, where the transmit power per LU is (a) 0 dBm and (b) -7 dBm, respectively.}
	\label{ResFig3l}
\end{figure}

Fig.~\ref{ResFig3l} illustrates the relationship between the ergodic rate per LU and the number of phase shift quantization bits.
We can see that the rate of the proposed IRS-enhanced anti-jamming precoder increases only slightly when the number of quantization bits is greater than 3.
The difference between the rate uusing a 2-bit versus a 3-bit IRS is also marginal. 
To offer a good compromise between the ergodic rate per LU and the complexity of solving $({\rm P} 4)$, 
we suggest that using a legitimate IRS with 2-bit phase shift quantization is sufficient.
\section{Conclusions}\label{Conclus} 
In this paper, an IRS-enhanced anti-jamming precoder was proposed. 
Since the AP can not access
the CSI of the DIRS-related channels and acquire knowledge of the time-varying
DIRS phase shifts in real-world applications, we maximized the received signal power by using the RCG algorithm to
design the passive beamforming at the IRS. Moreover, based on our investigations, 
using a 2-bit IRS provided a good trade-off between computational complexity and performance.
Compared with the previous work in~\cite{MyGC23,MyGC23Extension}, the proposed IRS-enhanced anti-jamming precoder 
can better mitigate the impact of DIRS-based ACA interference at high transmit power. 
In particular, the simulations showed that using an IRS that is one-seventh the size of the DIRS can essentially suppress
the impact of DIRS-based fully-passive jamming at 0 dBm transmit power.

\appendices
\section{Proof of Proposition~\ref{Proposition1}}\label{AppendixA}
According to~\eqref{Ricianchan} and~\eqref{Hdkeq},  
${\left[ {\bf H}_{\rm \!D}\!(t) \right]_{k,n}}$  
can be rewritten as
\begin{alignat}{1}
    \nonumber
    {\left[ {\bf H}_{\rm \!D}\!(t) \right]_{k,n}} &=  
    \sqrt {\frac{{{\varepsilon _n}{{{\mathscr{L}}}_{\rm{A\!D}}}{{\mathscr{L}}_{{\rm{D\!U}},k}}}}{{{\varepsilon _n} + 1}}} {\widehat {{\boldsymbol{h}}}_{{\rm{D\!U}},k}} \!\odot\! {\boldsymbol{\varphi}}\!_{\rm D}\!(t ) 
    \left[{\widehat {\bf{H}}}_{\rm{A\!D}}^{{\rm{LOS}}}\right]_{:,n} \\
    & + \sqrt {\frac{{{{{\mathscr{L}}}_{\rm{A\!D}}}{{\mathscr{L}}_{{\rm{D\!U}},k}}}}{{{\varepsilon _n} + 1}}} {\widehat {{\boldsymbol{h}}}_{{\rm{D\!U}},k}} \!\odot\! {\boldsymbol{\varphi}}\!_{\rm D}\!(t ) 
    \left[{\widehat {\bf{H}}}_{\rm{A\!D}}^{{\rm{NLOS}}}\right]_{:,n},
    \label{RewriHACAele}
\end{alignat}
where  $\odot$ represents the Hadamard product.
Furthermore, 
\begin{alignat}{1}
    \nonumber
    & {\left[  {\bf H}_{\rm \!D}\!(t) \right]_{k,n}}  \!=\! \\ \nonumber
    & \;\;\;\;\; \sqrt {\!\frac{ {{\varepsilon _n}{{{\mathscr{L}}}_{\rm{\!A\!D}}}{{\mathscr{L}}_{{\rm{\!D\!U}},k}}}}{{{\varepsilon _n} \!+\! 1}}}\! \sum\limits_{r = 1}^{{N_{\rm{\!D}}}} \!\!\left(\!{\!\left[\!{\widehat {{\boldsymbol{h}}}_{{\rm{D\!U}},k}}\!\right]_{r} \!\!
    {\!e^{j{\varphi _{{\rm \!D},r}}\!({t})}}\!\!\left[\!{\widehat {\bf{H}}}_{\rm{\!A\!D}}^{{\rm{LOS}}}\!\right]_{r,n}}\!\right) \\
      &\;\;\;\;\;  +\sqrt {\!\frac{\!{{{{\mathscr{L}}}_{\rm{\!A\!D}}}{{\mathscr{L}}_{{\rm{\!D\!U}},k}}}}{{{\varepsilon _n} \!+\! 1}}} \!\sum\limits_{r = 1}^{{N_{\rm{\!D}}}} \!\!\left(\!{\!\left[\!{\widehat {{\boldsymbol{h}}}_{{\rm{D\!U}},k}}\!\right]_{r} \!\!
    {e^{j{\varphi _{{\rm \!D},r}}\!({t})}}\!\!\left[\!{\widehat {\bf{H}}}_{\rm{A\!D}}^{{\rm{NLOS}}}\!\right]_{r,n}}\!\right).
    \label{Rewri18HACAele1} 
\end{alignat}
Conditioned on the fact that the random variables in~\eqref{Rewri18HACAele1} are independent, we have the following expectations
\begin{alignat}{1}
    &{\mathbb{E}}\!\!\left[{ \left[{\widehat {{\boldsymbol{h}}}_{{\rm{D\!U}},k}}\right]_{r} \!\!
    {e^{j{\varphi _{{\rm D},r}}\!({t})}}\!\!\left[{\widehat {\bf{H}}}_{\rm{A\!D}}^{{\rm{LOS}}}\right]_{r,n}} \right] 
    = 0, \label{Expect1} \\
    & {\mathbb{E}}\!\!\left[ { \left[{\widehat {{\boldsymbol{h}}}_{{\rm{D\!U}},k}}\right]_{r} \!\!
    {e^{j{\varphi _{{\rm D},r}}\!({t})}}\!\!\left[{\widehat {\bf{H}}}_{\rm{A\!D}}^{{\rm{NLOS}}}\right]_{r,n}} \right] = 0 .
    \label{Expect2}
\end{alignat}
Moreover, the variances of $ ({  [{\widehat {{\boldsymbol{h}}}_{{\rm{D\!U}},k}} ]_{r} 
{e^{j{\varphi _{{\rm D},r}}\!({t})}}\!  [{\widehat {\bf{H}}}_{\rm{A\!D}}^{{\rm{LOS}}} ]_{r,n}} )$ and
$ ({  [{\widehat {{\boldsymbol{h}}}_{{\rm{D\!U}},k}} ]_{r}  
{e^{j{\varphi _{{\rm D},r}}\!({t})}}\!  [{\widehat {\bf{H}}}_{\rm{A\!D}}^{{\rm{NLOS}}} ]_{r,n}} )$ are 
calculated as
\begin{alignat}{1}
    &{\rm{Var}}\!\!\left[{ \left[{\widehat {{\boldsymbol{h}}}_{{\rm{D\!U}},k}}\right]_{r} \!\!
    {e^{j{\varphi _{{\rm D},r}}\!({t})}}\!\!\left[{\widehat {\bf{H}}}_{\rm{A\!D}}^{{\rm{LOS}}}\right]_{r,n}} \right] 
    = 1, \label{Var1} \\
    & {\rm{Var}}\!\!\left[ { \left[{\widehat {{\boldsymbol{h}}}_{{\rm{D\!U}},k}}\right]_{r} \!\!
    {e^{j{\varphi _{{\rm D},r}}\!({t})}}\!\!\left[{\widehat {\bf{H}}}_{\rm{A\!D}}^{{\rm{NLOS}}}\right]_{r,n}} \right] = 1 .
    \label{Var2}
\end{alignat}

Based on the Lindeberg-L$\acute{e}$vy central limit theorem, 
the elements ${\left[ {\bf H}_{\rm \!D}\!(t) \right]_{k,n}}$
converge in distribution to a Gaussian distribution with zero mean and variance ${\beta _k}$ when the number of the DIRS reflective elements is large enough, i.e.,
\begin{equation}
    {\left[ {\bf H}_{\rm \!D}\!(t) \right]_{k,n}} \mathop  \to \limits^{\rm{d}} \mathcal{CN}\!\left( {0,  {\beta _k} } \right), {\rm{as}}\;N_{\rm \!D} \to \infty,
    \label{HDSta}
\end{equation}
where  
${\beta _k} = {{{\mathscr{L}}\!_{{\rm G}}}{{\mathscr{L}}\!_{{\rm I},k}}{N\!_{\rm D}} }$.
Substituting~\eqref{HDSta} to~\eqref{eqSLNR}, we can reduce ${\eta _k}$ to the form in~\eqref{eqSLNRredadd}.

\end{document}